\documentclass[prl,amsmath,aps,twocolumn]{revtex4}
\usepackage{graphicx,xspace}
\usepackage{float}
\usepackage{amsmath}
\usepackage{amssymb}
\usepackage[normalem]{ulem}
\input{epsf}
\begin{document}
\def\tit#1#2#3#4#5{{\sl #5} #1 {\bf #2} #3 (#4)}
\newcommand{\VEC}[1]{\mathbf{#1}}
\newcommand{\TO}{,\ldots,} 
\newcommand{\rvec}{\VEC{r}}
\newcommand{\scalar}[2]{\left \langle#1\ #2\right \rangle}
\newcommand{\me}{\mathrm{e}}
\newcommand{\mi}{\mathrm{i}}
\newcommand{\dif}{\mathrm{d}}
\newcommand{\ket}{\rangle}
\newcommand{\fcs}{f_{\text{c}}^{\text{smpl}}}
\newcommand{\bra}{\langle}
\newcommand{\half}{\frac{1}{2}}
\newcommand{\mean}[1]{\langle #1 \rangle}
\newcommand{\fig}[1]{figure~\ref{#1}}
\newcommand{\msmall}{\scriptscriptstyle}
\newcommand{\eq}[1]{eq.~\ref{#1}}
\newcommand{\gl}{\!=\!}
\newcommand{\mn}{\!-\!}
\newcommand{\nufs}{\nu_{\msmall \text{FS}}}
\newcommand{\pf}{\!\rightarrow\!}
\newcommand{\mq}[2]{\uwave{#1}\marginpar{#2}} 
\newcommand{\gae}{\lower 2pt \hbox{$\, \buildrel {\scriptstyle >}\over {\scriptstyle
\sim}\,$}}
\newcommand{\lae}{\lower 2pt \hbox{$\, \buildrel {\scriptstyle <}\over {\scriptstyle
\sim}\,$}}

\title{ 
Critical exponents of the driven elastic string in a disordered medium}
\author{Olaf Duemmer}
\email{duemmer@lps.ens.fr}
\author{Werner Krauth}
\email{krauth@lps.ens.fr}
\homepage{http://www.lps.ens.fr/~krauth}
\affiliation{CNRS-Laboratoire de Physique Statistique \\
Ecole Normale Sup{\'{e}}rieure\\
24 rue Lhomond, 75231 Paris Cedex 05, France}
\begin{abstract} 
We analyze the harmonic elastic string driven through a continuous random
potential above the depinning threshold. The velocity exponent $\beta =
0.33(2)$ is calculated. We observe a crossover in the roughness exponent
$\zeta$ from the critical value $1.26$ to the asymptotic (large force)
value of $0.5$.  We calculate directly the velocity correlation function
and the corresponding correlation length exponent $\nu = 1.29(5)$,
which obeys the scaling relation $\nu = 1/(2-\zeta)$, and agrees
with $\nufs$, the finite-size-scaling exponent of fluctuations in the
critical force. The velocity correlation function is non-universal at
short distances.
\end{abstract}
\maketitle

Driven elastic manifolds in disordered media model
the physics of systems as diverse as charge density waves \cite{Gruener},
interfaces in disordered magnets \cite{Lemerle}, contact lines of liquid
menisci on rough substrates \cite{Prevost}, vortices in type-II
superconductors \cite{Blatter} and crack propagation in solids \cite{Gao}.

An elastic manifold, driven through disorder by an external force,
undergoes a dynamic phase transition \cite{Fisher} that arises from
competition between the driving force and the pinning energy due to
the disorder, mediated by the elasticity of the manifold. Analogous to
an equilibrium phase transition, the driving force acts as the control
parameter, and the average center-of-mass velocity $v$ of the manifold
acts as order parameter.  Two phases of respectively zero and non-zero
order parameter are separated by the critical force $f_c$.  At forces
below this depinning threshold $f_c$, the disorder `pins' the manifold and
for long enough times, the velocity of the manifold is zero.  Above $f_c$,
the manifold continues to advance in avalanches. When the threshold
is approached from above ($f \pf f_c^+$), the mean velocity tends
to zero, and typical length, width and duration of the avalanches
diverge. This critical divergence is characterized by two independent
scaling exponents.

Much effort \cite{Dong,NarayanFisher,Chauve,Nattermann} has been spent
on calculating the universal exponents, particularly for driven elastic
manifolds in the limit of quasi-static motion. In this limit, inertial
terms are neglected and the force acting on the manifold is assumed
independent of velocity. The net force on the manifold comprises a
constant driving force $f$, a position-dependent random force $\eta$
and an elastic restoring force. Choosing a harmonic short range elastic
force leads to the following equation of motion for the manifold $h$
at zero temperature
\begin{equation}
\partial_t h(x,t) 
=f + \eta[x,h(x,t))] + \partial^2_x h(x,t).
\label{eq_motion}
\end{equation} 
In general, the manifold is represented as a single-valued function
$h(x,t)$ defined over a $D$-dimensional transversal space $x$, moving in
a $D+1$ dimensional disorder. In this paper, we consider one discrete
dimension $D=1$ with periodic boundary conditions for the string
of length $L$ ($x+L$ is identified with $x$) and for the disorder of
lateral extension $M$ ($h+M$ is identified with $h$, up to a winding term).
The continuous disorder of unit strength and unit correlation range is
constructed like in \cite{AlbertoRough}.

In what follows, we numerically investigate the dynamics of the
$1+1$-dimensional string above the depinning threshold. Convergence of
the dynamical solution is ensured by exploiting the particular analytical
structure \cite{Middleton,Baesens} of the equation of motion: it has a
unique periodic solution for each disorder sample in the $t \pf \infty$
limit. Having found the periodic solution, we are thus
certain to have reached the asymptotic regime and to have shaken off all
influence of arbitrary initial conditions.  The asymptotic periodic solution is
constructed to desired precision using a continuous integration routine
for \eq{eq_motion}. Averaging observables over one period and over
disorder samples, we calculate the velocity exponent $\beta$ and describe
in detail how the correlation length diverges as $f \pf f_c^+$.
\begin{figure}
\centerline{\includegraphics{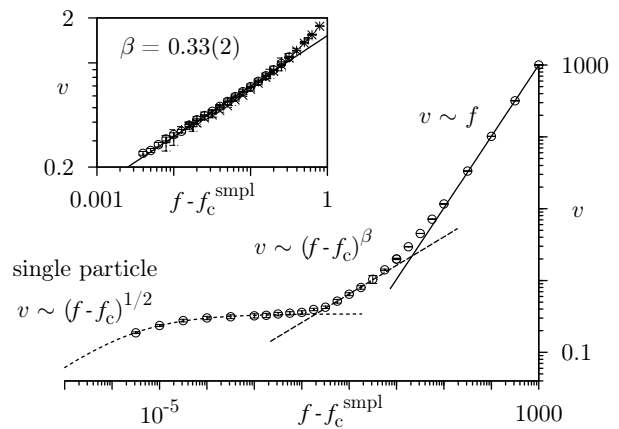}}
\caption{Mean velocity $v$ of the elastic string vs
  $f-\fcs$, ($L=128, M=195$, $18$ samples).  The window of critical
  power law behavior is sandwiched between the single-particle regime 
  $v \sim f^{1/2}$ at very small $f \!-\! \fcs$ --- fitted to eq.\ref{eq_fit}
  --- and linear $v \sim f$ behavior at large $f$.  Inset: critical window for larger system sizes
  $L=512 \TO 2048$ with $M\sim L^{\zeta}$. Slope on log-log plot yields
  the velocity exponent $\beta = 0.33(2)$.}
\label{fig_beta}
\end{figure}

In the thermodynamic limit, the string velocity obeys a power law:
$v \sim (f-f_c)^{\beta}$ for $f \rightarrow f_c$.  On finite systems,
the critical force $\fcs$ fluctuates from sample to sample, putting
a limit on how small the control parameter $f \mn \mean{\fcs}$ can be
made without introducing undesired corrections to scaling relations;
a limit from which previous numerical calculations suffer.

We are able to determine the exact sample-dependent depinning threshold
$\fcs$ thanks to a recent algorithm \cite{AlbertoAlgo}.  The sample
critical force itself shows non-negligible fluctuations of the order
of $\sigma_{f_c} \approx L^{-1/\nufs}$.  The algorithm also finds the
final critical configuration $h_c$ and the roughness exponent $\zeta$
at depinning \cite{AlbertoRough}.

Knowing the critical force $\fcs$ of each sample, we plot the
time- and disorder-averaged velocity $v$ against $f-\fcs$.
Thus we eliminate the statistical noise due to the fluctuating
critical force, and obtain extraordinarily clean data. Furthermore, the
control parameter $f \mn \fcs$ can be made arbitrarily small, which is
not possible when using $f-\mean{\fcs}$.

The mean velocity on a \emph{finite} sample shows three different
regimes (see \fig{fig_beta}): For very small $f-\fcs$, the motion of the
entire string is correlated, and it behaves effectively like a single
particle ($v \sim (f-\fcs)^{1/2}$). At intermediate forces, the string
is correlated on length-scales $\xi$ and moves collectively ($v \sim
(f-\fcs)^{\beta}$). At high forces the motion is essentially uncorrelated
($v \sim f$).

At $f=\fcs$, the finite dynamical system (\eq{eq_motion}) (in the $t \pf
\infty$  limit) undergoes a saddle node bifurcation between a static
(pinned) and a periodic (depinned) solution; the global velocity
minimum changes from stable to unstable. For very small positive
$f-\fcs$, the string spends the major part of its time-period passing
through the velocity minimum, and negligibly little time completing its
orbit. The motion through the minimum is dominated by the mode $\tilde
h$ with the largest eigenvalue in a linear stability analysis around
the critical configuration $h_c$. The mode $\tilde h$ moves at a velocity equal to
$f-\fcs$ plus quadratic and higher corrections in $\tilde h$. Hence we
can express the time spent inside the minimum as: $\int \frac {\dif h}
v = \int \frac {\dif h}{f-\fcs + c{\tilde h}^2} = T_{\text{i}} (f-\fcs)^{-1/2}$,
to first order in $f-\fcs$. The remaining time $T_{\text{o}}$ spent outside the
velocity minimum depends only weakly on $f-\fcs$. This yields for the velocity
as a function of force: 
\begin{equation}
\label{eq_fit}
  M/v = T = T_{\text{o}} + T_{\text i}(f-\fcs)^{-1/2}. 
\end{equation}
As shown in \fig{fig_beta}, this function
--- combining the effective single-particle exponent one half ($ v \sim
(f- \fcs)^{1/2}$) for very small forces with the saturation at slightly
larger forces --- fits the data perfectly.

At intermediate forces, the dynamics leave the single-particle regime and
enter the regime of critical collective motion.  The cross-over takes
place when the dynamic correlation length $\xi$, which diverges as
$\xi \sim (f-f_c)^{-\nu}$, equals the system size $L$ \cite{Cardy}. It
follows that the difference between the corresponding cross-over force
and the sample critical force scales as $L^{-1/\nu}$. This scaling
is quite slow because of the small exponent $1/\nu \approx 0.7$, as our
data confirms. The other cross-over,  between the critical region and the
linear regime, is independent of system size.

Between the two cross-overs lies the window of collective behavior,
small and slowly growing with $L$: Only strings of length $L \gtrsim 512$
allow to see any significant evidence for critical behavior, and even
for  $L=2048$ (see inset of \fig{fig_beta}) the window is less
than two decades.  These pronounced finite-sample-size effects limiting
the critical window (also observed in the CDW model \cite{Myers}) could
obscure critical behavior in experimental situations. For example, in
an experiment of a liquid-solid contact line advancing on a disordered
substrate \cite{Moulinet} the sample size, i.e. the capillary length
acting as upper cutoff, is $L\approx 200$ in units of the disorder
correlation length. For these small samples we expect the window
of collective behavior to be hardly noticable and finite-size effects
to dominate.

When extrapolating data to the thermodynamic limit, we have to carefully
choose the lateral sample size: $M$ has to be of the order of the
typical width $W$ of a string of length $L$ at depinning. When increasing $L$,
$W$ scales as $L^{\zeta}$, $\zeta$ being the roughness exponent. Hence the
lateral sample size $M$ has to scale as $L^{\zeta}$, too. Otherwise, if $M$ is scaled with an exponent
$\zeta' < \zeta$, the periodic sample is too short, the string wraps around
it, and generates correlations which mix in properties of the charge density
wave model (CDW, $M \sim 1$), itself member of a different universality
class \cite{NarayanFisherCDW}. If on the other hand $M$ scales with $\zeta'
> \zeta$, the sample is too long and contains $\sim M/W \sim L^{\zeta'-\zeta} $ independent
critical configurations of size $L \times W$. Each of these
configurations has a slightly different {\em local} critical force, and
the critical force of the entire sample is given by their maximum, and
not their mean. Consequently the critical force of a sample of size $M \gg W$
overestimates $f_c$, even in the thermodynamic limit.

We therefore use a scaling of $M\sim L^{\zeta}$, with strings of length
$L = 512,...,2048$, when analyzing the mean velocity.  Inside the window
of critical collective behavior, we find $\beta = 0.33(2)$ (see inset
of figure \ref{fig_beta}). This value is consistent with $0.2(2)$ from
a two-loop functional renormalization \cite{Chauve},
and with $0.35(5)$ from a simulation of an advancing magnetic domain
wall in a 2D random Ising system \cite{Nowak}. It is larger than previous
estimates on continuous systems $0.22(2)$ \cite{Zapperi} and on automaton
models $0.25(3)$ \cite{Nattermann}. In these studies, however, $v$ was
plotted against $f-\mean{\fcs}$, and $\beta$ was probably measured partly
inside the critical window and  partly inside the finite-size-effect
region where the velocity saturates. This naturally leads to $\beta$ 
being underestimated.

\begin{figure}
\centerline{\includegraphics{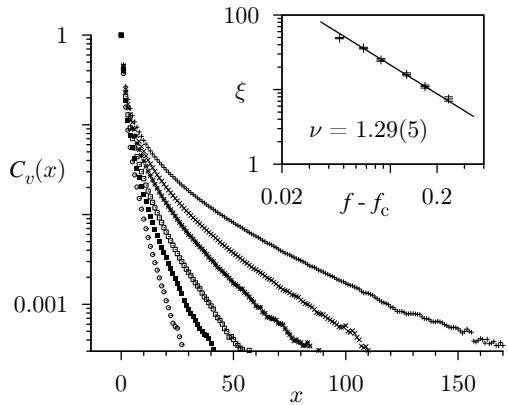}}
\caption{Connected velocity-velocity correlations $C_v(x)$ for different
  values of $f-f_c$, ($L=512$, $M=20000$, $700$ samples).
  Inset: correlation length $\xi$ as obtained from a fit to
  $ C_v(x) = C_0 x^{-\kappa} e^{-x/ \xi }$.}
\label{fig_vv}
\end{figure}

At the  depinning threshold the typical size $\xi$ of avalanches diverges.
Whereas previous work did not succeed in accessing $\xi$ directly,
we are able to compute this dynamic correlation length from
the velocity correlation function and, independently, the structure
factor.  In our continuous calculation, contrary to automaton models,
the instantaneous velocities $v(x,t)$ are readily available, and the
velocity correlation function can be computed unequivocally:
\begin{equation*}
C_v(x) = \bra v(x)v(0) \ket_c = \sum_{x'}^L \bra (v(x'+x,t)-\bar
v)(v(x',t)-\bar  v) \ket_t.
\end{equation*}

Two points on the string taking part in the same avalanche have
correlated
instantaneous velocities. Therefore, the typical length
$\xi$ of avalanches shows up as the characteristic length scale in the
velocity correlation function.

Analyzing this correlation function $C_v(x)$ we find a diverging
correlation length $\xi$ (see \fig{fig_vv}). In order to extract $\xi$,
we perform for every given $f-f_c$ a two-parameter ($\xi$,$\kappa$)
fit of the correlation function to the  functional form
\begin{equation}
C_v(x) \simeq C_0 \cdot x^{-\kappa}\me^{-x / \xi}.
\label{eq_cv}
\end{equation}

The inset of \fig{fig_vv} shows the correlation length $\xi$ obtained
as a function of $f-f_c$. It diverges as $\xi \sim (f-f_c)^{-\nu}$
with $\nu = 1.29(5)$. This value agrees with $\nu_{ \text{FS}} =
1.33(1)$ found in a finite-size-scaling analysis of sample-to-sample
fluctuations of the critical force \cite{AlbertoFluct}.  Hence we confirm
that the dynamic correlation length exponent and the finite-size-scaling
correlation length exponent are identical $\nu = \nu_{\msmall \text{FS}}$
\cite{NarayanFisher,Chauve}. Furthermore, $\nu$ obeys the statistical
tilt symmetry relation $\nu = 1/(2-\zeta)$ \cite{Nattermann}.

The velocity correlation function is found to be non-universal: the
exponent $\kappa$ characterizing the algebraic decay at small distances
depends on $f-f_c$, unlike previously thought \cite{NarayanFisher,Dong}.
This is why the correlation function $C_v(x)$ for different values of $f - f_c$
cannot be plotted on one single master curve as a function of
$x/\xi$. We have not attempted to establish a definite functional relation
between $\kappa$ and $f-f_c$.

The non-universal behavior of the velocity correlation function at
small distances might be linked to the fact that in one dimension the
harmonic model's  roughness exponent is greater than one: The average 
distance between neighboring points grows without
bound, implying that any real string described by this model would eventually
break \cite{AlbertoPRL}. Preliminary data on a model with higher than
harmonic elasticity \cite{DuemmerKrauth} does indicate universal velocity
correlations, suggesting that the harmonic short range model is indeed
exceptional.

\begin{figure}
\centerline{\includegraphics{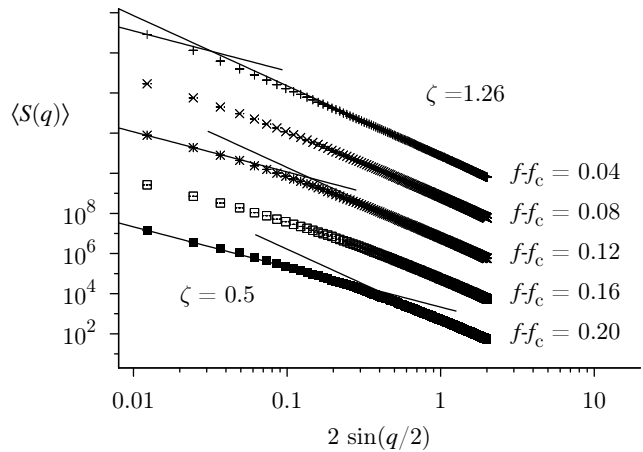}}
\caption{
  The structure factor $\mean{S(q)}\sim 1/q^{1+2\zeta}$
  for different values of $f - f_c$ ($L=512, M=20000$, $200$ samples). Two
  regimes of different roughness $\zeta$ are visible:
  At small $q$-values $\zeta = 1/2$, while for large $q$, $\zeta = 1.26$. The
  crossover tends toward zero for $f \rightarrow f_c$, illustrating the
  diverging correlation length. Curves for different $f-f_c$ are offset along
  the y-axis, lines are guides to the eye.}
\label{fig_rough}
\end{figure}

An avalanche of typical length $\xi$ has a typical width $w$ which scales
like $w \sim \xi^{\zeta}$. As a second method to estimate the roughness
exponent $\zeta$, we study the time- and disorder-averaged structure
factor $S(q)$, which behaves like $1/q^{\,1+2\zeta}$, when defined as
$S(q) = \bra h(q)h(-q) \ket_c$ with $h(q) = \sum_x^L \me^{\mi qx}h(x)$.

At large driving forces $f \rightarrow \infty$, the roughness exponent
of the harmonic elastic string is $\zeta=1/2$: The noise due to the disorder
becomes equivalent to thermal noise, which can be seen from expanding
the disorder term $\eta$ in eq. \ref{eq_motion} of motion --
transformed to the center of mass reference frame -- in powers of
$v^{-1} \approx f^{-1}$, yielding a $\delta(t)$ correlated noise to first
order. At depinning, the roughness exponent takes on the value $\zeta =
1.26(1)$ \cite{AlbertoRough}.

Inside the critical window, both these values for $\zeta$ show up in the
structure factor, and characterize two different ranges of $q$ separated
by the inverse of the correlation length: for large wave-vectors ($q
\gtrsim \frac{2\pi}{\xi}$) the structure factor displays the critical roughness,
whereas at small wave-vectors ($q \lesssim \frac{2\pi}{\xi}$) it shows the thermal
roughness. Figure \ref{fig_rough} tracks the crossover
between the two regimes as  $f \rightarrow f_c^+$, illustrating the
diverging correlation length $\xi$. The corresponding correlation length
exponent $\nu$ is consistent with the value calculated from the velocity
correlation function, but less precise.

In our calculation of the correlation function, the lateral size $M$
must be chosen sufficiently large for the string not to notice the
periodicity of the disorder --- otherwise, the structure factor displays
at small $q$, in addition to the two regimes mentioned, the roughness
of a CDW ($\zeta_{\msmall \text{CDW}} = 3/2$ \cite{NarayanFisherCDW}).
This means that the smallest $q$-modes have to completely decorrelate on
the scale $M$.  The mode $q$ decorrelates in a time  $\sim q^{-z}$ with
$z$ the dynamic exponent. $z$ not only controls the typical duration of
the avalanches (of size $\xi(f-f_c)$), but also the decorrelation time
of \emph{all} length scales \cite{foot_convergence}, in particular the 
string length $L$. Note that inside the critical window $L > \xi$.  
This implies that at a mean
velocity of order unity, i.e. at the upper end of the critical window,
the smallest mode needs a distance of roughly $L^z$ to decorrelate. But
$z\approx 1.5$ is larger than $\zeta$.  We therefore have to choose an
$M \gg L^{\zeta}$ in our analysis of $S(q)$ and $C_v(x)$, in order to
eliminate auto-correlation effects in the largest length-scales.

On the contrary, in our initial analysis of the mean velocity (see
\fig{fig_beta}), we were justified in adopting a scaling $M\sim L^{\zeta}$
because the effective string viscosity --- which determines $\beta$ ---
is caused by pinning on length scales \emph{below} $\xi$.
The periodicity  $M \sim L^{\zeta}$ thus does not influence the
velocity exponent $\beta$.

At very large $M$, the  sample critical force $\fcs$ becomes too large to
be used as the critical force.  We therefore calculate a mean critical
force $f_c^{\infty}$ from a finite-size-scaling ansatz: The average
sample critical force is taken to depend on the sample length $L$ as
$f_c^{\infty} - \mean{\fcs(L)} \sim L^{-1/\nu_{\msmall \text{FS}}}$. The
asymptotic value $f_c^{\infty} = 1.913(2)$ is then used as a mean critical
force when analyzing the velocity correlations and the structure factor.

When investigating the correlation length $\xi$ directly by means of
$C_v(x)$ and $S(q)$, we do not need to know $\fcs$ in order to avoid the
cross-over from the critical window into the finite-size-effect region:
We simply limit our analysis to sufficiently small values of $\xi/L$. In
addition, the smallest value of $f-f_c^{\infty}$ analyzed is larger than
typical fluctuations in $\fcs$, which are therefore negligible.

In conclusion, we analyze the quasi-static dynamics of the harmonic
elastic string driven through a random potential, above the depinning
transition. We calculate the exact periodic solution for each sample, and
encounter large finite-sample-size effects, which will have repercussions
in experimental situations. Knowing exactly the sample critical force
enables us to identify the limited window of critical collective behavior
and to determine the velocity exponent $\beta = 0.33(2)$.  We investigate
the velocity correlation function, determine the correlation length
exponent $\nu = 1.29(5)$, and confirm that it both obeys the statistical
tilt symmetry and agrees with the finite-size-scaling correlation length
exponent.  Surprisingly, we find a non-universal functional form for
the velocity correlation function: the exponent $\kappa$ describing the
short distance behavior depends on the control parameter $f-f_c$.

We would like to thank P. Le Doussal, S. Guibert, A. Rosso and K. Wiese  for
stimulating discussions.


\begin{thebibliography}{10}
\bibitem{foot_convergence} In fact, $z$ governs the convergence of our
 numerical algorithm (cf \cite{DuemmerKrauth}).
\end{thebibliography}
\end{document}